\documentclass{elsart}

\usepackage{graphicx}
\usepackage{epsfig}

\usepackage{amssymb}

\newcommand\fverb{\setbox\pippobox=\hbox\bgroup\verb}
\newcommand\fverbdo{\egroup\medskip\noindent%
            \fbox{\unhbox\pippobox}\ }
\newcommand\fverbit{\egroup\item[\fbox{\unhbox\pippobox}]}
\newbox\pippobox
\def\ff{$\phi$--factory}
\def\ifm#1{\relax\ifmmode#1\else$#1$\fi}  \def\DAF{DA\char8NE}  \def\x{\ifm{\times}}
\def\pt#1,#2,{\ifm{#1\x10^{#2}}}   \def\up#1{\ifm{^{#1}}}   \def\dn#1{\ifm{_{#1}}}
\def\plm{\ifm{\,\pm}\,}   \def\ab{\ifm{\sim}}   \def\deg{\ifm{^\circ}}   
  \def\to{\ifm{\rightarrow}}   \def\f{\ifm{\phi}}  
\def\ks{\ifm{K_S}}  \def\kpm{\ifm{K^\pm}}  \def\kp{\ifm{K^+}}  \def\km{\ifm{K^-}}
         \def\epm{\ifm{e^+e^-}}
\def\kb{\ifm{\rlap{\kern.3em\raise1.9ex\hbox to.6em{\hrulefill}} K}}

  \def\po{\ifm{\pi^0}}    \def\rmk{\rm\kern.5mm }
   \def\L{\ifm{{\cal L}}}
\def\figb#1;#2;{\parbox{#2cm}{\epsfig{file=#1.eps,width=#2cm}}}
\def\figbc#1;#2;{\cl{\figb #1;#2;}}    \def\dt{\ifm{{\rm d}\,t}}
\def\ie{{\it\kern-1pt i.\kern-.5pt e.\kern-.2pt}}  
\let\cl=\centerline
\catcode`@=11 
\newdimen\z@ \z@=0pt 
\newskip\z@skip \z@skip=0pt plus0pt minus0pt
\def\m@th{\mathsurround=\z@}
\def\ialign{\everycr{}\tabskip\z@skip\halign} 
\def\eqalign#1{\null\,\vcenter{\openup\jot\m@th
  \ialign{\strut\hfil$\displaystyle{##}$&$\displaystyle{{}##}$\hfil
      \crcr#1\crcr}}\,}
\catcode`@=12 

\newcommand{\BR}[1]{\ensuremath{{\rm BR}(#1)}}

\newcommand{\eq}[1]{eq.~\ref{#1}}

\newcommand{\Tab}[1]{Table~\ref{#1}}

\newcommand{\kpmunug} {\ensuremath{K^{+}\rightarrow \mu^{+}\nu~\gamma}}
\newcommand{\kpmunugall} {\ensuremath{K^{+}\rightarrow \mu^{+}\nu~(\gamma)}}
\newcommand{\kpmunu} {\ensuremath{K^{+}\rightarrow \mu^{+}\nu}}

\newcommand{\kmmunugall} {K^{-}\rightarrow \mu^{-}\kern.3em\overline{\nu}\,(\gamma)}

\let\kpppg=\kppipigall
\begin{document}

\begin{frontmatter}

\title{Measurement of the absolute branching ratio of the $K^{+} \rightarrow \pi^{+}\pi^0\,(\gamma)$ decay with the KLOE detector}

\collab{The KLOE Collaboration}
\author[Na,infnNa]{F.~Ambrosino},
\author[Frascati]{A.~Antonelli},
\author[Frascati]{M.~Antonelli},
\author[Roma2,infnRoma2]{F.~Archilli},
\author[Roma3,infnRoma3]{C.~Bacci},
\author[Karlsruhe]{P.~Beltrame},
\author[Frascati]{G.~Bencivenni},
\author[Frascati]{S.~Bertolucci},
\author[Roma1,infnRoma1]{C.~Bini},
\author[Frascati]{C.~Bloise},
\author[Roma3,infnRoma3]{S.~Bocchetta},
\author[Frascati]{F.~Bossi},
\author[infnRoma3]{P.~Branchini},
\author[Frascati]{P.~Campana},
\author[Frascati]{G.~Capon},
\author[Frascati]{T.~Capussela},
\author[Roma3,infnRoma3]{F.~Ceradini},
\author[Roma3,infnRoma3]{F.~Cesario},
\author[Frascati]{S.~Chi},
\author[Na,infnNa]{G.~Chiefari},
\author[Frascati]{P.~Ciambrone},
\author[Roma1]{F.~Crucianelli},
\author[Frascati]{E.~De~Lucia\corauthref{cor}}
\ead{erika.delucia@lnf.infn.it}
\corauth[cor]{Corresponding author.}
\author[Roma1,infnRoma1]{A.~De~Santis},
\author[Frascati]{P.~De~Simone},
\author[Roma1,infnRoma1]{G.~De~Zorzi},
\author[Karlsruhe]{A.~Denig},
\author[Roma1,infnRoma1]{A.~Di~Domenico},
\author[infnNa]{C.~Di~Donato},
\author[Roma3,infnRoma3]{B.~Di~Micco},
\author[infnNa]{A.~Doria},
\author[Frascati]{M.~Dreucci},
\author[Frascati]{G.~Felici},
\author[Frascati]{A.~Ferrari},
\author[Frascati]{M.L.~Ferrer},
\author[Roma1,infnRoma1]{S.~Fiore},
\author[Frascati]{C.~Forti},
\author[Roma1,infnRoma1]{P.~Franzini},
\author[Frascati]{C.~Gatti},
\author[Roma1,infnRoma1]{P.~Gauzzi},
\author[Frascati]{S.~Giovannella},
\author[Lecce,infnLecce]{E.~Gorini},
\author[infnRoma3]{E.~Graziani},
\author[Karlsruhe]{W.~Kluge},
\author[Moscow]{V.~Kulikov},
\author[Roma1,infnRoma1]{F.~Lacava},
\author[Frascati]{G.~Lanfranchi},
\author[Frascati,StonyBrook]{J.~Lee-Franzini},
\author[Karlsruhe]{D.~Leone},
\author[Frascati,Moscow]{M.~Martemianov},
\author[Frascati,Energ]{M.~Martini},
\author[Na,infnNa]{P.~Massarotti},
\author[Frascati]{W.~Mei},
\author[Na,infnNa]{S.~Meola},
\author[Frascati]{S.~Miscetti},
\author[Frascati]{M.~Moulson},
\author[Frascati]{S.~M\"uller},
\author[Frascati]{F.~Murtas},
\author[Na,infnNa]{M.~Napolitano},
\author[Roma3,infnRoma3]{F.~Nguyen},
\author[Frascati]{M.~Palutan},
\author[infnRoma1]{E.~Pasqualucci},
\author[infnRoma3]{A.~Passeri},
\author[Frascati,Energ]{V.~Patera},
\author[Na,infnNa]{F.~Perfetto},
\author[infnLecce]{M.~Primavera},
\author[Frascati]{P.~Santangelo},
\author[Na,infnNa]{G.~Saracino},
\author[Frascati]{B.~Sciascia},
\author[Frascati,Energ]{A.~Sciubba},
\author[Frascati]{A.~Sibidanov},
\author[Frascati]{T.~Spadaro},
\author[Roma1,infnRoma1]{M.~Testa},
\author[infnRoma3]{L.~Tortora},
\author[infnRoma1]{P.~Valente},
\author[Frascati]{G.~Venanzoni},
\author[Frascati,Energ]{R.~Versaci},
\author[Frascati,Beijing]{G.~Xu}
\address[Frascati]{Laboratori Nazionali di Frascati dell'INFN, Via E. Fermi 40, I-00044 Frascati, Italy.}
\address[Karlsruhe]{Institut f\"ur Experimentelle Kernphysik, Universit\"at Karlsruhe, D-76128 Karlsruhe,Germany.}
\address[Lecce]{Dipartimento di Fisica dell'Universit\`a del Salento, Via Arnesano, I-73100 Lecce, Italy.}
\address[infnLecce]{INFN Sezione di Lecce, Via Arnesano, I-73100 Lecce, Italy.}
\address[Na]{Dipartimento di Scienze Fisiche dell'Universit\`a  di Napoli ``Federico II'', Via Cintia, I-80126 Napoli, Italy.}
\address[infnNa]{INFN Sezione di Napoli, Via Cintia, I-80126 Napoli, Italy.}
\address[Energ]{Dipartimento di Energetica della  Sapienza Universit\`a di Roma, P. Aldo Moro 2, I-00185 Roma, Italy.}
\address[Roma1]{Dipartimento di Fisica della Sapienza Universit\`a di Roma, P. Aldo Moro 2, I-00185 Roma, Italy.}
\address[infnRoma1]{INFN Sezione di Roma, P. Aldo Moro 2, I-00185 Roma, Italy.}
\address[Roma2]{Dipartimento di Fisica dell'Universit\`a di Roma ``Tor Vergata'', 
Via della Ricerca Scientifica 1, I-00133 Roma, Italy.}
\address[infnRoma2]{INFN Sezione di Roma Tor Vergata, Via della Ricerca Scientifica 1, I-00133 Roma, Italy.}
\address[Roma3]{Dipartimento di Fisica dell'Universit\`a di Roma ``Roma Tre'', Via della Vasca Navale 84, I-00146 Roma, Italy.}
\address[infnRoma3]{INFN Sezione di Roma Tre, Via della Vasca Navale 84, I-00146 Roma, Italy.}
\address[StonyBrook]{Physics Department, State University of New York at Stony Brook, Stony Brook, NY 11794-3840  USA.}
\address[Beijing]{Institute of High Energy Physics of Academia Sinica,  P.O. Box 918 Beijing 100049, P.R. China.}
\address[Moscow]{Institute for Theoretical and Experimental Physics, B. Cheremushkinskaya ul. 25 RU-117218 Moscow, Russia.}
\begin{abstract}
We have measured the absolute branching ratio of the \kpppg\ decay, using \ab20 million tagged $K^{+}$ 
mesons collected with the KLOE detector at \DAF, the Frascati \ff.
Signal counts are obtained from the fit of the distribution
of the momentum of the charged decay particle in the kaon rest frame. 
The result, inclusive of final-state radiation, is \BR{\kpppg}=0.2065\plm0.0005\dn{\rm stat}\plm 0.0008\dn{\rm syst}.
\end{abstract}
\end{frontmatter}
\section{Introduction}
The branching ratio of the \kpppg\ decay ($K_{\pi2}$)
is part of the KLOE program of precise and fully
inclusive kaon branching ratios (BRs) measurement.
We have already measured the main $K_{L}$~\cite{klbr,klpipi} and $K_{S}$~\cite{ks2pi,kslept}
branching ratios.
We report here our measurement of the \BR{\kpppg} which together with $\BR{K^{\pm}\to\mu^{\pm}\nu}$~\cite{kmu2},
$\BR{K^{\pm}\to\pi^0 l^{\pm}\nu}$~\cite{kpmkl3} and $\BR{K^{\pm}\to\pi^{\pm}\pi^0\pi^0}$~\cite{plbtauprime} covers
95\% of all charged kaon decays.
The importance of this measurement is twofold:
i) the most recent measurement  
 based on 16,000 events from a sample of \ab10\up5 kaon decays, \BR{\kpppg}=0.2118\plm0.0028~\cite{chiang}~, 
dates back to more than 30 years ago and gives no information on the radiation cut-off and
ii) this BR is necessary to obtain $\BR{K_{\rm l3}}$ from measurements normalized to $\BR{K_{\pi2}}$~\cite{na48,istra+}.
The \kpppg\ branching ratio can be used, together with the $\ks\to\pi\pi$ branching ratios~\cite{ks2pi}, to determine the
relative phase $\delta_0 - \delta_2$ of the $I$=0 and $I$=2 s-wave $\pi\pi$-scattering amplitudes~\cite{cirigliano:kppiso}.
In the following we report our measurement of the absolute branching ratio \BR{\kpppg}
performed with the KLOE detector using an integrated luminosity $\int\!\L\dt\,$\ab250 pb\up{-1} collected 
at \DAF , the Frascati \ff . 
\DAF\ is an \epm\ collider operated at the energy of 1020 MeV, the mass of the
\f\ meson. Equal energy positron and electron beams collide at an
angle of ($\pi-$0.025) radians producing \f-mesons with a transverse momentum of \ab13 MeV.
In its rest frame, the \f-meson decays into anti-collinear
\kp\km\ pairs of \ab127 MeV momentum and this remains approximately true in the laboratory.
Detection of a $K^\pm$ (the tagging kaon) therefore signals the presence of a $K^\mp$ (the tagged kaon) 
of given momentum and direction. This procedure, called tagging, allows measurements of absolute BRs.
\section{The KLOE detector}
\label{sec:detector}
The KLOE detector consists of a large volume drift chamber surrounded by an electromagnetic sampling calorimeter. The entire detector is immersed in an axial magnetic field $B = 0.52$ T.
The drift chamber (DC) \cite{DC}, 3.3 m long and 4 m in
diameter, has a stereo geometry with 12,582 drift cells arranged in 58 layers
and operates with a 90\% helium-10\%
isobutane gas mixture. Tracking in the DC provides measurements of the
 momentum of charged particles with
$\sigma(p_{\perp})/p_{\perp}\leq0.4\%$ for polar angles larger than 45$^{\deg}$. The spatial resolution is 
\ab150 $\mu$m in the bending plane, \ab2 mm on the $z$ coordinate and \ab3 mm on decay vertices.
The electromagnetic calorimeter (EMC) \cite{EMC} consists of a cylindrical
barrel and two endcaps, covering a solid angle of 98\% of 4$\pi$. 
Particles crossing the lead-scintillator-fiber structure of the EMC, 
segmented into five planes in depth, are detected as local energy deposits.
Deposits close in time and space are grouped into clusters.
The energy and time resolution for electromagnetic showers
are $\sigma_{E}/E = 5.7\%/\sqrt{E(\mbox{GeV})}$ and
$\sigma_{t} = 57\mbox{ ps}/\sqrt{E(\mbox{GeV})} \oplus 100 \mbox{ ps}$, respectively.
The trigger \cite{TRIG} requires two isolated energy
deposits in the EMC with: E~$>$~50 MeV in the barrel and E~$>$~150 MeV in the endcaps. 
Cosmic-ray muons are identified as events with two energy
deposits with E~$>$~30 MeV in the outermost EMC planes and vetoed at the trigger level (CRV). 
A software filter (SF), based on the topology and multiplicity
of EMC clusters and DC hits, is applied
to reject machine background. 
The effect of both CRV and SF on the BR measurement must be determined.
In the following the coordinate system is defined with the
$z$-axis along the bisector of the \epm\ beams, the $y$-axis vertical
and the $x$-axis toward the center of the collider rings and origin at the collision point.
\section{The measurement}
%
%
Tagging with $K^-\to\mu^-\nu$ ($K^{-}_{\mu2}$) and
$K^-\to\pi^-\pi^0$ ($K^{-}_{\pi2}$) decays provides two samples
of pure $K^+$ for signal search.
These two-body decays are easily identified as peaks in the distribution of the 
$p^{\ast}_{\pi}$ variable, the momentum of the charged decay particle in the kaon rest frame evaluated
using the pion mass, as described in \cite{kmu2_KN205}.
The tagging kaon is required to satisfy
the trigger request by itself,
minimizing the dependence of the trigger efficiency on the decay mode of the tagged kaon.
The residual dependency, which we refer to as the tag bias in the following,
 must be determined for the BR evaluation.
We choose to measure $\BR{K_{\pi 2}}$ using \kp\ mesons because for them
the nuclear interaction correction is negligible, 
since the probability of interaction is \ab10$^{-5}$ for \kp\ and \ab3.4\% for \km.
%
%

The branching ratio is determined as:
\begin{eqnarray}
   {\rm BR}\ (\kpppg) =
   \frac{N_{\kpppg}}{N_{\rm Tag}}\times {1\over\:\epsilon\:C_{\rm CRV}\:C_{\rm SF}\:C_{\rm TB}}
   \label{eq:brformula}
\end{eqnarray}
where $N_{\kpppg}$ is the signal count, $N_{\rm Tag}$ the number of tagged events and $\epsilon$ is 
the overall efficiency, including the detector acceptance $\epsilon_{\rm det}$ and
the reconstruction efficiency $\epsilon_{\rm rec}$.
The detector acceptance ($\epsilon_{\rm det}$\ab59\%), entering in the final efficiency evaluation,
is taken from MC and its value is related to the charged kaon lifetime $\tau$.
Consequently the BR depends on $\tau$ as:
\begin{equation}
 {\rm BR}^{(\tau)}/{\rm BR}^{(0)}=1-0.0395 {\rm ~ns}^{-1}(\tau-\tau^{(0)})
\label{eq:brvstau}
\end{equation}
with $\tau^{(0)}=12.385$ ns, the current world average value~\cite{pdg06up}. 
A variation of the lifetime of 0.1\% changes the BR of 0.05\% of its value.
The corrections $C_{\rm CRV}$, $C_{\rm SF}$ and $C_{\rm TB}$ account for
the cosmic-ray muons veto, the software filter and tag bias effects, respectively.

The sample used for this measurement has been 
processed and filtered with the KLOE standard reconstruction software
and event classification procedure \cite{OFFLINE}.
The KLOE Monte Carlo (MC) simulation package, GEANFI, has been used to produce an event sample equivalent to the data.
The different operating conditions of \DAF\ during data taking,
 machine parameters and background, are included in the MC on a run-by-run basis.
%
%
The simulation also includes final-state radiation~\cite{mcgen} 
guaranteeing correct measurement of fully inclusive BRs.
The result of the simulation has been compared
with theoretical predictions and experimental results for several kaon decay channels.
 
\subsection{$K^{-}_{\mu2}$-tagged sample}
\label{sec:kmu2_sample}
The number of $K^+$ tagged by $K^{-}_{\mu2}$ decays, the $K^{-}_{\mu2}$-tagged sample,
 is $N_{\rm Tag}$ = 12,113,686.
The $K^{+}_{\pi2}$ signal selection uses DC information only.
The \kp\ track is identified as a positive track with point of closest approach (PCA) to the interaction point (IP) satisfying 
$\sqrt{x^2_{\rm PCA} + y^2_{\rm PCA}}<$ 10 cm and $|z_{\rm PCA}|<$  20 cm, and momentum 70 $<p_{\rm K}<$
130 MeV. The PCA is evaluated extrapolating the $K^{+}$ track 
backwards to the IP taking into account
energy losses. Decay vertices (V) are accepted in the fiducial volume 
40 $< \sqrt{x^2_{\rm v} + y^2_{\rm v}}<$ 150 cm, $|z_{\rm v}|<$150 cm.
Loose cuts on $p^{\ast}_{\pi}$ and on the difference between the momenta of the kaon and the charged secondary track,
$50 < p^{\ast}_{\pi} < 370$ MeV and $-320 < \Delta p < -50 $ MeV, reject $K\to 3 \pi$ decays.
%
%
%
\begin{figure}
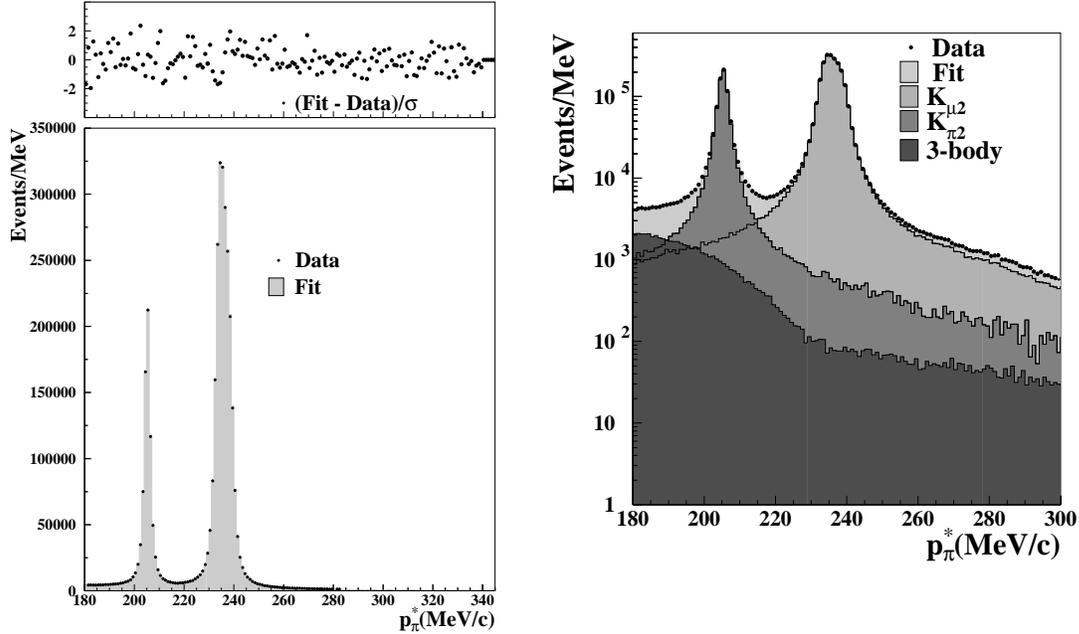

\figb fig1a;7.;\kern0.2cm \figb fig1b;7.;
\caption{Fit of the $p^{\ast}_{\pi}$ distribution. Left: black dots are data and
grey histogram is the fit output. Right: the three contributions used to fit the data
are shown: $K_{\mu 2}$, $K_{\pi 2}$ and three-body decays.
\label{fig:fit_res}}
\end{figure}

The $K^{+}_{\pi2}$ signal count is extracted  from the fit of the $p^{\ast}_{\pi}$ distribution
 (Fig.~\ref{fig:fit_res}).
The $p^{\ast}_{\pi}$ spectrum has two peaks: the first at \ab236 MeV due to muons from $K_{\mu2}$ decays,
and the second at \ab205 MeV due to pions from $K_{\pi2}$ decays.
The contribution from three-body decays shows at lower $p^{\ast}_{\pi}$ values.
Having used the pion mass for the $p^{\ast}_{\pi}$ evaluation, 
the $K_{\mu2}$ peak is distorted.
We fit the $p^{\ast}_{\pi}$ distribution between 180 and 350 MeV using 
three contributions: $K_{\mu 2}$, $K_{\pi 2}$
and three-body decays.
The shapes of the $K_{\mu 2}$ and $K_{\pi 2}$ peaks are obtained from data control samples,
selected using EMC information only.
The $K_{\pi 2}$ spectrum is obtained from the $K_{\pi 2}$-control-sample
 used for the efficiency evaluation and described later.
The  $K_{\mu 2}$ spectrum
is obtained from the control sample selected for the \BR{\kpmunu} measurement~\cite{kmu2}.
Once a tagging $K^-_{\mu2}$ decay has been identified, 
we ask for only one EMC cluster with energy $E_{\rm Clu}>$~80 MeV and no clusters with energy between 
20 and 80 MeV. There are no requirements on EMC clusters with energy below 20 MeV, in order 
to retain \kpmunug\ decays and \kpmunu\ decays with machine background clusters in the EMC. 
This high-purity sample (\ab99\%) is called $K_{\mu 2}$-control-sample.
Bin by bin MC corrections account for small
distortions induced in the $p^{\ast}_{\pi}$ spectra by the control sample selections.
The three-body component is obtained from the MC simulation, which has been tuned with the data 
$K_{\mu 2}$- and $K_{\pi 2}$-control-sample.
Fig.~\ref{fig:fit_res} left shows the result of the fit of the
$p^\ast_{\pi}$ distribution compared to the data, while 
the three different contributions are visible on the right.
The fit,
with  $\chi^2/{\rm ndf}=154.5/156$ (P$(\chi^2)=$0.52),
 gives $N_{\kpppg}=818,347 \pm 1,912$, the error accounting for the
 statistics (not only data).

%
%

The reconstruction efficiency $\epsilon_{\rm rec}$  has been evaluated with data. Since the $K^+_{\pi2}$ events 
are identified from DC information, the data control sample is selected using EMC information.
Once a tagging $K^-_{\mu2}$ decay has been identified, we construct by kinematics the \kp\ 
track from the \km\ track. We then search for two photons in the EMC and, using their time and energy 
information, we determine the \kp\ decay point, the di-photon mass and momentum. The best accuracy
is obtained minimizing the sum of the square of the differences between the decay time from 
photons and \kp\ path
and between the di-photon mass and the \po\ mass. Having determined the track and decay point of the \kp\ and 
the \po\ direction,
we determine the expected $\pi^{+}$ track using the two-body decay hypothesis. The kinematics of this hypothesis is then 
verified by requiring the presence of a cluster in the EMC, with a distance from the pion track $d_{\rm Clu} < $ 30 cm.
These events define the $K_{\pi 2}$-control-sample. 
The contamination from \kp\ decays without a $\pi^0$ in the final state is $\sim 0.1\%$.
 About 5\% contamination from $K_{l3}$ decays is present and becomes about 3\% after signal selection.
Corrections
accounting for small distortions due to the selection of the data
control sample have been evaluated using MC.
Defining $\epsilon_{\rm true}$ the true efficiency to reconstruct signal decays in the DC volume
and $\epsilon_{\rm cs}$ the reconstruction efficiency obtained using the $K_{\pi2}$-control-sample,
the average correction to be applied to the efficiency is $\epsilon_{\rm true}/ \epsilon_{\rm cs}\sim 0.99$.
The efficiency to be used in \eq{eq:brformula} is $\epsilon = 0.3176 \pm 0.0005$.
\begin{table}[!h!b!t]
\begin{center}
\begin{tabular}{cc}\hline\hline
\multicolumn{2}{c}{Statistical errors using $K^-_{\mu 2}$ tag}\\
\hline Source & Value (\%)\\
\hline fit signal count & 0.23 \\
efficiency & 0.12 \\
SF and CRV & 0.04 \\
$\epsilon$ correction & 0.12 \\ 
TB & 0.05 \\ \hline
Total &  0.30 \\ \hline\hline\end{tabular}
\kern1cm
\begin{tabular}{cc}\hline\hline
\multicolumn{2}{c}{Statistical errors using $K^-_{\pi 2}$ tag}\\
\hline Source  & Value (\%) \\ 
\hline fit signal count & 0.27 \\
efficiency & 0.13 \\
SF and CRV & 0.03 \\
$\epsilon$ correction & 0.13 \\
TB & 0.05 \\ \hline
Total & 0.33\\ \hline\hline\end{tabular}
\caption{Summary of fractional statistical uncertainties on $\BR{K^+_{\pi 2}}$ measured using 
$K^-_{\mu 2}$- and $K^-_{\pi 2}$-tagged samples. Left $K^-_{\mu 2}$ tag, right $K^-_{\pi 2}$ tag.}
\label{tab:stat}
\end{center}
\end{table}

The corrections $C_{\rm CRV}=1.0005 \pm 0.0003$ and
$C_{\rm SF}= 1.0183\pm0.0003$ have been measured with
data taken without the cosmic-ray muons veto and the software filter, respectively.
The correction for the tag bias, $C_{\rm TB} = 1.0106 \pm 0.0005_{\rm stat}$,
has been evaluated using MC. The distributions of 
variables used for the selection of the tagging decay have been checked with data.
\Tab{tab:stat} left lists the statistical fractional uncertainties on the branching ratio measurement
and the total value is 0.3\%.
\subsection{$K^{-}_{\pi2}$-tagged sample}
\label{sec:kpi2_sample}
The number of $K^+$ tagged by $K^{-}_{\pi2}$ decays, the $K^{-}_{\pi2}$-tagged sample,
 is $N_{\rm Tag}=9,352,915$.
\Tab{tab:br_corr} compares the values of the $C_{\rm CRV}$, $C_{\rm SF}$ and $C_{\rm TB}$ corrections obtained for 
$K^-_{\mu 2}$- and $K^-_{\pi 2}$-tagged events.
\begin{table}[!h!b!t]
\begin{center}
\begin{tabular}{ccc}
\hline
\hline
& $K^-_{\mu 2}$ tag & $K^-_{\pi 2}$ tag \\
\hline
$C_{\rm CRV}$   & 1.0005$\pm$0.0003 &  1.0007$\pm$0.0003 \\
$C_{\rm SF}$ & 1.0183$\pm$0.0003 &  1.00093$\pm$0.00006 \\
$C_{\rm TB}$ & 1.0106$\pm$0.0005 &  1.009$\pm$0.0006 \\
\hline\hline
\end{tabular}
\caption{Corrections to $\BR{K^+_{\pi 2}}$ measured using 
$K^-_{\mu 2}$- and $K^-_{\pi 2}$-tagged samples.}
\label{tab:br_corr}
\end{center}
\end{table}
The two tags have very different corrections for the effect of the software filter (SF).
The $C_{\rm SF}$ correction measured using the $K^-_{\mu 2}$ tag is \ab1.8\% 
while using the $K^-_{\pi 2}$ tag is \ab0.1\%.
The same signal selection as before is applied to the sample tagged by
$K^{-}_{\pi2}$ decays and the fit of the $p^\ast_{\pi}$ distribution determines
the signal count.
The spectra used for the fit have been obtained as described
in the previous section, once a tagging $K^-_{\pi 2}$ decay has been identified.
The signal count is $N_{\kpppg}=621,612\pm1,678$.
For the efficiency evaluation we have used the $K_{\pi 2}$-control-sample tagged
by $K^{-}_{\mu2}$ decays. The efficiency is $\epsilon = 0.3182 \pm 0.0005$,
 corrected for the control sample selection and the detector acceptance taken from MC.
The total statistical fractional uncertainty on $\BR{K^-_{\pi 2}}$ measured
using the $K^{-}_{\pi2}$-tagged sample is 0.33\%.
\Tab{tab:stat} right summarizes the fractional statistical uncertainties.
\section{Systematic uncertainties}
The systematic uncertainties on  $\BR{K^+_{\pi2}}$ from $K^{-}_{\mu2}$ and $K^{-}_{\pi2}$-tagged samples are listed in tables \ref{tab:syst} left and right, respectively.
The stability of both BR measurements with respect to different data taking periods and conditions
has been checked. A detailed discussion of the systematic studies follows.
These studies have been done varying the selection cuts in wide intervals and checking the stability of the BR.
\begin{table}
\begin{center}
\begin{tabular}{cc}\hline\hline
\multicolumn{2}{c}{Systematic errors using $K^-_{\mu 2}$ tag}\\\hline
Source & Value (\%) \\\hline
$p^\ast_{\pi}$ fit range & 0.06 \\$K_{\mu 2}$ shape & 0.12 \\
$K_{\pi 2}$ shape & 0.16 \\ efficiency & 0.30 \\
$\rho^{\rm min}_{\rm v}$ & 0.17 \\ lifetime $\tau$ & 0.12  \\
TB & 0.01 \\ Nucl. int. &  $<$ 0.02 \\
\hline Total & 0.42 \\\hline\hline\end{tabular}\kern1cm
\begin{tabular}{cc}\hline\hline
\multicolumn{2}{c}{Systematic errors using $K^-_{\pi 2}$ tag}\\\hline
Source  & Value (\%) \\\hline
$p^\ast_{\pi}$ fit range & 0.07 \\ $K_{\mu 2}$ shape & 0.14 \\
$K_{\pi 2}$ shape & 0.17 \\ efficiency & 0.30 \\
$\rho^{\min }_{\rm v}$ & 0.17 \\ lifetime $\tau$ & 0.12 \\
TB & 0.01 \\Nucl. int. & $<$ 0.02 \\\hline
Total & 0.43 \\\hline\hline\end{tabular}
\caption{Summary of fractional systematic uncertainties on $\BR{K^+_{\pi 2}}$ measured using 
$K^-_{\mu 2}$- and $K^-_{\pi 2}$-tagged samples. Left $K^-_{\mu 2}$ tag, right $K^-_{\pi 2}$ tag.}
\label{tab:syst}
\end{center}
\end{table}

The lower bound of the $p^\ast_{\pi}$ fit range, 180 MeV,
has been moved from 165 to 190 MeV, changing by almost a factor of two the
contribution from three-body decays. For each value of the lower bound, we have performed the fit of the $p^\ast_{\pi}$
distribution and evaluated the overall efficiency.
We observe a minimal change of the BR value in the above range.
The maximum variation of the BR is taken as systematic uncertainty. The contributions to the
fractional systematic uncertainty on the BR are 0.06\% ($K^-_{\mu2}$ tag) and 0.07\% ($K^-_{\pi2}$ tag).

The spectrum of the $K_{\mu 2}$ component for the
fit of the $p^\ast_{\pi}$  distribution
 is obtained from the $K_{\mu 2}$-control-sample.
The $K_{\mu 2}$ spectrum is most affected by the cut at 20 MeV
on the cluster energy $E_{\rm Clu}$ \cite{kmu2_KN205}, 
connected to the acceptance of a photon from $\kpmunug$ decays
or from machine background events.
The stability of the BR measurement has been checked by changing the 
$E_{\rm Clu}$ cut from 10 to 30 MeV, corresponding to a change in the purity of the
$K_{\mu 2}$-control-sample from \ab99.3\% to \ab97\%. Negligible effects are observed with $E_{\rm Clu}$
values larger than 30 MeV. 
The maximum variation of the BR has been taken as systematic uncertainty. 
The fractional systematic uncertainties are 0.12\% ($K^-_{\mu2}$ tag) and 0.14\% ($K^-_{\pi2}$ tag).

The spectrum of the $K_{\pi 2}$ component for the fit of the $p^\ast_{\pi}$
distribution is obtained from the $K_{\pi 2}$-control-sample.
The systematic effect has been estimated performing the fit
with the $p^\ast_{\pi}$ spectrum obtained from a different control sample.
We select \kp\ decays in the DC, using the signal selection
of sec.~\ref{sec:kmu2_sample}, and require the identification of a $\pi^0$,
 looking for two photons in the EMC fulfilling the following requests.
The two photons have to be on-time: the difference between the kaon decay times,
evaluated using the cluster time and the distance between the \kp\ decay vertex and 
the cluster position, has to be within 3$\sigma_{\rm t}$ (see sec.~\ref{sec:detector}). 
The kaon decay time from the kaon path and from the photons have to be compatible within resolutions.
 The difference between the di-photon mass and the $\pi^0$ mass has to be within 3$\sigma$, with $\sigma$\ab18 MeV.
The $K_{l3}$ contamination of this sample is 20\%, larger than the 3\% contamination of the $K_{\pi 2}$-control-sample.
Thus the spectrum of the $K_{\pi 2}$ component obtained from this sample needs larger MC bin by bin corrections (as large as 60\%)
compared to the default used
(20\% at maximum and for low values of $p^\ast_{\pi}$).
Using this spectrum
we have been performed the fit of the $p^\ast_{\pi}$ distribution, also varying the fit range. 
The BR results are in agreement, within errors, with the values obtained using the 
spectrum from the $K_{\pi 2}$-control-sample. 
The maximum difference between the BRs obtained with the two spectra has been taken as systematic uncertainty. 
The fractional contribution is 0.16\% ($K^-_{\mu2}$ tag) and 0.17\% ($K^-_{\pi2}$ tag).

The reconstruction efficiency  has been evaluated with data using the
$K_{\pi 2}$-control-sample. The systematic uncertainty has been estimated
using a different control sample, with larger $K_{l3}$ contamination (\ab11\% compared to \ab3\%) and
MC correction to be applied to the efficiency (\ab12\% compared to \ab1\%).
\kp\ decays with a $\pi^0$ in the final state are selected, as done for the $K_{\pi 2}$-control-sample but
without the $d_{\rm Clu}$ cut. We determine the $p^{\ast}_{\pi}$ of the charged secondary track, 
using the two-body hypothesis and 
the \kp\ and $\pi^0$ momenta. Two-body decays are then selected applying 
the asymmetric cut $0.5\sigma<p^{\ast}_{\pi}-205<\sigma$, with $\sigma$\ab18 MeV, around the peak 
 at 205 MeV of the $p^{\ast}_{\pi}$ distribution.
The BRs measured using the efficiencies obtained from the above sample and the 
$K_{\pi 2}$-control-sample agree within errors.
Conservatively the difference between these two BRs is taken 
as the systematic uncertainty. The contribution to the fractional systematic uncertainty is 0.3\%.

The \kp\ decay vertex has to satisfy the 
requirement 40 $< \rho_{\rm v}=\sqrt{x^2_{\rm v} + y^2_{\rm v}}<$ 150 cm.
The lower bound of the $\rho_{\rm v}$ range, $\rho^{\rm min}_{\rm v}=40$ cm, has been moved from 38 to 42 cm
with the detector acceptance changing of \ab6\% of its value.
For each $\rho^{\rm min}_{\rm v}$ value, we have performed the fit, evaluated the efficiency and measured
the BR. The efficiency has been evaluated with the $K_{\pi 2}$-control-sample. The resolution on the 
\kp\ decay point, using only the time information in the EMC, is $\sigma$\ab1.5 cm. 
Thus the above interval corresponds to a change of more than 2$\sigma$.  
The BR results are in agreement within the statistical error and their rms is taken as systematic uncertainty.
The contribution to the fractional systematic uncertainty is 0.17\%.

The BR depends on the charged kaon lifetime $\tau$ through the detector acceptance.
The systematic effect has been obtained
using \eq{eq:brvstau}
and the 0.24\% fractional accuracy
of the KLOE measurement $\tau$=12.347$\pm$0.030 ns~\cite{tau_kpm}.
The contribution to the fractional systematic uncertainty is 0.12\%.

The fractional systematic uncertainty from the tag
definition is 0.01\%, as obtained changing separately
the requirements to identify the tagging decay.

The fraction of $K^+$ undergoing nuclear interaction has been evaluated
using the MC simulation and considered as upper bound value of the
systematic uncertainty. The contribution to the fractional systematic uncertainty 
 is $<$0.02\%.

The emission of radiation in the 
\ensuremath{K^{+} \rightarrow \pi^{+}\pi^0\gamma} decay is dominated by the
Inner Bremsstrahlung (IB) contribution and the Direct Emission term can be
neglected~\cite{adler}. The predicted value of the IB branching ratio is 
2.61$\times$10$^{-4}$ and our simulation reproduces this value within few
10$^{-6}$. Thus, the systematic uncertainty associated to the modelling
of final-state radiation is negligible.
The average of the experimental results is
BR(\ensuremath{K^{+} \rightarrow \pi^{+}\pi^0\gamma}) =
(2.75$\pm$0.15)$\times$10$^{-4}$~\cite{pdg06up}
and its error gives a negligible contribution to the uncertainty on $\BR{K^+_{\pi2}}$.

The fit of the $p^\ast_{\pi}$ distribution providing the count
for $K^+_{\pi2}$ decays, gives the number of $\kpmunugall$ decays as well.
The reliability of the fit procedure is confirmed by comparing $\BR{K^+_{\mu2}}$
and finding agreement with our published result~\cite{kmu2}.
The criteria for signal selection and efficiency
evaluation from this reference have been followed. 
There is therefore a correlation $\rho(K_{\mu 2},K_{\pi 2})$ of $-$3.4\%
 between our $\BR{K^+_{\pi2}}$ and $\BR{K^+_{\mu2}}$ measurements using the
signal count extracted from the fit procedure. 
For our published $\BR{K^+_{\mu2}}$ result~\cite{kmu2} we did not use the fit of the $p^\ast$ distribution to extract
the signal count. The number of $\kpmunugall$ decays was obtained by counting the number of events with $p^{\ast}>$ 225 MeV,
 after background subtraction. Therefore there is no correlation between the 
published $\BR{K^+_{\mu2}}$ value and our $\BR{K^+_{\pi2}}$ value.

When averaging the $\BR{K^+_{\pi2}}$ values obtained from the $K^{-}_{\mu2}$-
and $K^{-}_{\pi2}$-tagged samples we have to account for correlations.
The same data control sample for efficiency evaluation has been used for both measurements,
thus giving a correlation in the statistical as well as in the systematic
contribution to the BR uncertainty. The contribution to the systematic uncertainty 
from the charged kaon lifetime value $\tau$ is common to both measurements
 as well as the contribution from the $\rho^{\rm min}_{\rm v}$ value.
The correlation between the two $\BR{K^+_{\pi2}}$ measurement is 56\%.
\section{Conclusions}
We have measured the branching ratio of the \kpppg\ decay,
fully inclusive of final-state radiation, using \kp\ samples tagged by $K^{-}_{\mu2}$
and $K^{-}_{\pi2}$ decays.
From 12,113,686 $K^{-}_{\mu2}$-tagged events, 
we find $N_{\kpppg}=818,347\plm1,912$ signal counts.
Using eq.~\ref{eq:brformula} we obtain the branching ratio:
\begin{equation}
  {\rm BR} \left.(K^+ \rightarrow \pi^+ \pi^0 (\gamma))\right|_{\rm K_{\mu2}-tag} =
   0.20638 \pm 0.00062_{\rm stat} \pm 0.00087_{\rm syst}.
   \label{eq:br_kpi2}
\end{equation}
From 9,352,915 $K^{-}_{\pi2}$-tagged events we have
$N_{\kpppg}=621,612\plm1,678$ signal counts corresponding to:
\begin{equation}
   {\rm BR}  \left.(K^+ \rightarrow \pi^+ \pi^0 (\gamma))\right|_{\rm K_{\pi2}-tag} =
   0.20668 \pm 0.00068_{\rm stat} \pm 0.00089_{\rm syst}.
   \label{eq:br_kpi2_tagpi2}
\end{equation}
The above BRs are evaluated using the current average value for the \kpm\ lifetime
$\tau^{(0)}=12.385$ ns (see eq.\ref{eq:brvstau}).
Averaging these two results, accounting for correlations, we obtain:
\begin{equation}
   {\rm BR} (K^+ \rightarrow \pi^+ \pi^0 (\gamma)) =
  0.2065 \pm 0.0005_{\rm stat} \pm 0.0008_{\rm syst}.
   \label{eq:ave_br_kpi2}
\end{equation}
This absolute branching ratio measurement is fully inclusive of final-state radiation and 
has a  0.46\% accuracy.
Our result is 1.3\% (\ab2$\sigma$) lower than the PDG fit~\cite{pdg06up}.
The global fit to all available charged kaon measurements~\cite{flavianet:brs} gives
${\rm BR}(K^+ \rightarrow \pi^+ \pi^0 (\gamma)) = 0.2064 \pm 0.0008$, in agreement with our result.

We fit the six largest \kpm\ BRs and the lifetime $\tau$ 
using our measurements of $\tau$~\cite{tau_kpm}, $\BR{K^+_{\pi2}}$
(\eq{eq:ave_br_kpi2}), 
$\BR{K^+_{\mu2}}$~\cite{kmu2}, $\BR{K^\pm_{\rm l3}}$~\cite{kpmkl3} and
$\BR{K^{\pm}\to\pi^{\pm}\pi^0\pi^0}$~\cite{plbtauprime}, with their dependence on $\tau$, together with 
$\BR{\kpm\to\pi^\pm\pi^+\pi^-}$ from the
PDG04 average\footnote{PDG '06 gives the result of their constrained fit but not the average of the data}
~\cite{pdg04}, 
with the sum of the BRs constrained to unity. 
The fit results, with  $\chi^2/{\rm ndf}=0.59/1$ (CL=44\%), 
are shown in table~\ref{tab:kpmbrfit} and confirm
the validity of our measurement (\eq{eq:ave_br_kpi2}), assuming the correctness of $\BR{\kpm\to\pi^\pm\pi^+\pi^-}$.
\begin{table}[!h!b!t]{
\begin{tabular}{llccccccc}
\hline\hline
Parameter & Value & \multicolumn{6}{c}{Correlation coefficients} \\
\hline
$\BR{K_{\mu2}}$          & 0.6376(12)       &         &         &         &         &         &            \\
$\BR{K_{\pi2}}$          & 0.2071(9)        & $+0.48$ &         &         &         &         &            \\
$\BR{\pi^\pm\pi^+\pi^-}$ & 0.0553(9)        & $-0.48$ & $+0.21$ &         &         &         &            \\
$\BR{K_{e3}}$            & 0.0498(5)        & $+0.37$ & $-0.13$ & $+0.16$ &         &         &            \\
$\BR{K_{\mu3}}$          & 0.0324(4)        & $+0.34$ & $-0.12$ & $+0.15$ & $+0.58$ &         &            \\
$\BR{\pi^\pm\pi^0\pi^0}$ & 0.01765(25)      & $-0.11$ & $+0.05$ & $-0.05$ & $+0.04$ & $+0.04$ &            \\
$\tau$ (ns)             & 12.344(29)      & $-0.15$ & $-0.21$ & $-0.07$ & $-0.06$ & $-0.05$ & $-0.015$   \\
\hline\hline
\end{tabular}
\caption{Results of the fit to \kpm\ BRs.}
\label{tab:kpmbrfit}}
\end{table}

We can also evaluate $\BR{\kpm\to\pi^\pm\pi^+\pi^-}$ by using our measurements of the above listed BRs and 
imposing the constraint $\sum \BR{\kpm\to f}=1$. With $\BR{K^+_{\mu2}}=0.63660\pm0.00175$, 
$\BR{K^{\pm}\to\pi^{\pm}\pi^0\pi^0}= 0.01763\pm0.00025$ and  
$\BR{K^+_{\pi2}}= 0.20681\pm0.00094$, $\BR{K^\pm_{e3}}= 0.04972\pm0.00053$ and $\BR{K^\pm_{\mu3}}=0.03237\pm0.00039$, 
evaluated at $\tau$ equal to our measured value 12.347$\pm$0.030 ns, 
we get $\BR{\kpm\to\pi^\pm\pi^+\pi^-}= 0.0568\pm0.0022$.
This result is in agreement with the PDG04 average $\BR{\kpm\to\pi^\pm\pi^+\pi^-}=0.0550\pm0.0010$~\cite{pdg04}.

%
%
The $K\to\pi\pi$ data can be used to extract the s-wave $\pi\pi$ scattering
phase shift difference $\delta_0 - \delta_2$ at $s=m^2_{\rm K}$. This has
been done by the FlaviaNet Kaon Working Group~\cite{flavianet:delta2}, 
using as experimental inputs the FlaviaNet world average value of the BR for
 \kpppg\ decay~\cite{flavianet:brs} and the KLOE measurement of the BRs 
for $\ks\to\pi^0\pi^0$ and $\ks\to\pi^+\pi^-(\gamma)$ decays~\cite{ks2pi}. 
Their result is $\delta_0 - \delta_2 = (57.5 \pm 3.4)^\circ$,
 accounting for strong and electromagnetic isospin breaking effects.
The same calculation has been done using our value of the BR for
 \kpppg\ decay from the fit to \kpm\ BRs (table \ref{tab:kpmbrfit}).
The result is $\delta_0 - \delta_2 = (57.6 \pm 3.4)^\circ$ and confirms
 the \ab2$\sigma$ discrepancy with the results from phenomenological analysis  
of $\pi\pi$ scattering amplitudes~\cite{flavianet:delta2}.

\section*{Acknowledgments}
We thank the DAFNE team for their efforts in maintaining low background running
conditions and their collaboration during all data-taking.
We want to thank our technical staff:
G.F.~Fortugno and F.~Sborzacchi for their dedicated work to ensure an
efficient operation of
the KLOE Computing Center;
M.~Anelli for his continuous support to the gas system and the safety of
the
detector;
A.~Balla, M.~Gatta, G.~Corradi and G.~Papalino for the maintenance of the
electronics;
M.~Santoni, G.~Paoluzzi and R.~Rosellini for the general support to the
detector;
C.~Piscitelli for his help during major maintenance periods.
This work was supported in part
by EURODAPHNE, contract FMRX-CT98-0169;
by the German Federal Ministry of Education and Research (BMBF) contract 06-KA-957;
by the German Research Foundation (DFG),
'Emmy Noether Programme', contracts DE839/1-4;
by INTAS, contracts 96-624, 99-37;
and by the EU Integrated Infrastructure
Initiative HadronPhysics Project under contract number
RII3-CT-2004-506078.
%
%

\end{document}